\documentclass{article}

\usepackage{multirow}

\usepackage[final,nonatbib]{neurips_2021}

\usepackage[utf8]{inputenc} 
\usepackage[T1]{fontenc}    
\usepackage{hyperref}       
\usepackage{url}            
\usepackage{booktabs}       
\usepackage{amsfonts}       
\usepackage{nicefrac}       
\usepackage{microtype}      
\usepackage{xcolor} 
\usepackage{tabularx}
\usepackage{graphicx}
\usepackage{bm}

\title{Benchmarking deep generative models for diverse antibody sequence design}

\author{%
  Igor Melnyk \\
  IBM Research AI\\
  \texttt{igor.melnyk@ibm.com} \\
  \And
  Payel Das \\
  IBM Research AI\\
  \texttt{daspa@us.ibm.com} \\
  \And
  Vijil Chenthamarakshan \\
  IBM Research AI\\
  \texttt{ecvijil@us.ibm.com} \\
  \And
  Aurelie Lozano \\
  IBM Research AI\\
  \texttt{aclozano@us.ibm.com} \\
}

\begin{document}

\maketitle

\begin{abstract}
  Computational protein design, i.e. inferring novel and diverse protein sequences consistent with a given structure, remains a major unsolved challenge. Recently, deep generative models that learn from sequences alone or from sequences and structures jointly have shown impressive performance on this task. However, those models appear limited in terms of modeling structural constraints, capturing enough sequence diversity, or both. Here we consider three recently proposed deep generative frameworks for protein design: (AR) the sequence-based autoregressive  generative model, (GVP) the precise structure-based graph neural network, and Fold2Seq that leverages a fuzzy and scale-free representation of a three-dimensional fold, while enforcing structure-to-sequence (and vice versa) consistency. We benchmark these models on the task of computational design of antibody sequences, which demand designing sequences with high diversity for functional implication. The Fold2Seq framework outperforms the two other baselines in terms of diversity of the designed sequences, while maintaining the typical fold.
\end{abstract}

\section{Introduction}
Antibodies and their functional domains play key roles in research, diagnostics, and therapeutics. Among them, an attractive class is comprised of nanobodies, which are  functional antibody domains with small size ($\sim$15 kDa) and high stability (Tm up to 90 C), and therefore are of increasing therapeutic interest \cite{bannas2017nanobodies}. Designing functional sequences typically  requires a combinatorial exploration of sequence space. To address this issue, one can   impose sequence and/or structural constraints to narrow the search space. 

Functionally diverse antibodies correspond to the highly stable immunoglobulin fold, a universal protein scaffold structure. Antigen binding specificity is largely determined by the sequence and structural diversity of the complementarity-determining regions (CDRs) that are displayed on canonical frameworks.  Among the CDRs, CDR3 contributes most of the sequence and length diversity. As a result, a big focus in computational antibody design has been on sampling diverse CDR3s.

Along this direction, recently deep generative models have been used for sampling  virtual sequences that were then prioritized for experimental validation. Most of those models are sequence-based that leveraged autoregressive neural nets such as LSTM. Such models have been successfully used for the entire functional antibody sequence design \cite{saka2021antibody}, as well as for designing only the CDRs given rest of the sequence as a context \cite{shin2021protein}.  Nevertheless, such model does not explicitly account for the 3D structural constraints associated with the characteristic immunoglobulin fold and may be limited in terms of capturing the broad landscape of CDRs. 

In this paper, we benchmark three recently proposed state-of-the-art deep generative models --  sequence-based, 3D structure-based, and  3D fold-based -- by comparing the generated  sequences of llama nanobodies. The structure-based and fold-based models generate full protein sequences by using a single representative 3D structure of the llama nanobody as an input, whereas the autoregressive sequence-based model designs CDR3 by conditioning on the  preceding germline framework-CDR1-CDR2 nanobody sequence. The structure-based model leverages the precise backbone coordinates, whereas the fold-based model uses a fuzzy description of the secondary structures elements in 3D as the input. Machine learning, physicochemical, bioinformatics, and structural metrics of the generated sequences from a single llama nanobody sequence/structure and their comparison with the natural llama nanobody repertoire suggest that generated sequences strongly vary in term of novelty, diversity, consistency, and coverage. 


\begin{figure}[!t]
\centering
\includegraphics[width=1.0\textwidth]{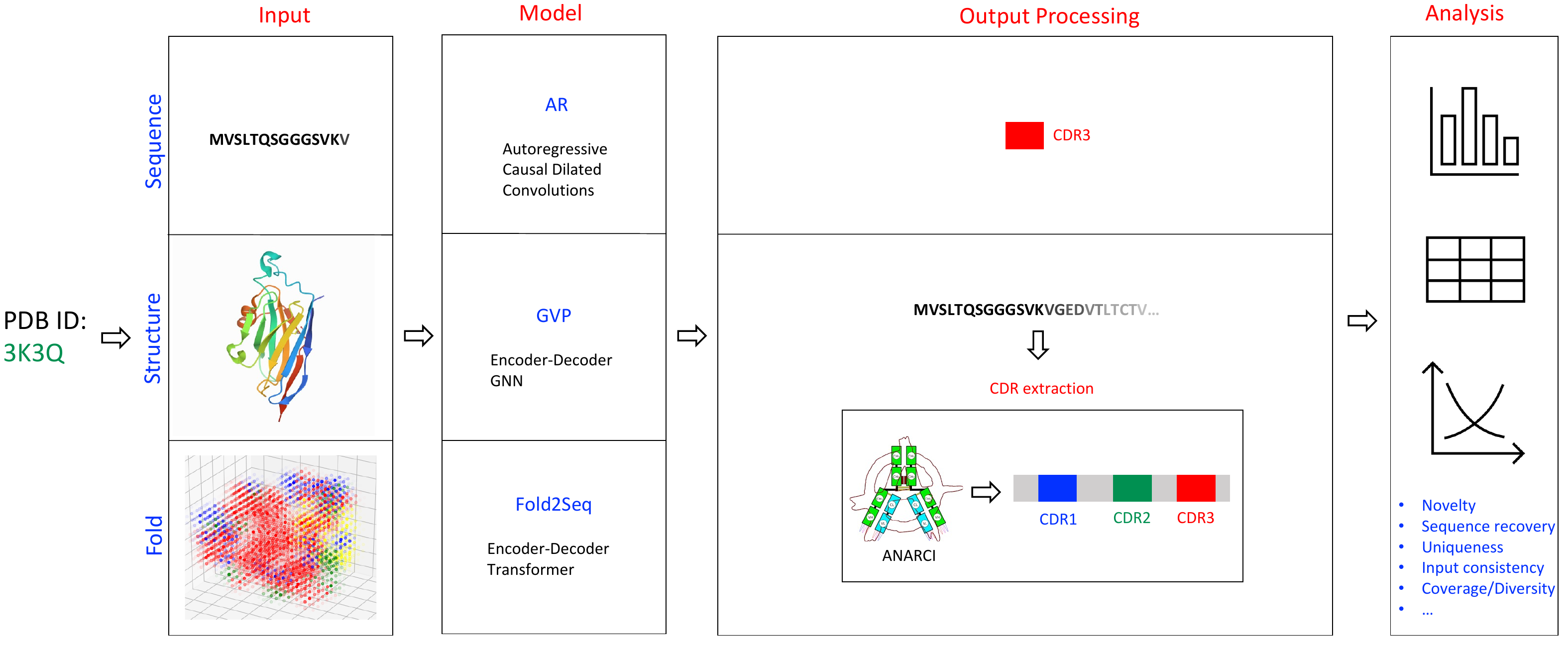}
\caption{Overview of our study for benchmarking models for antibody sequence design.  Starting with the prefix sequence (CDR1, CDR2 and the framework), structure or the fold of the PDB ID 3K3Q, we use three models i.e., AR \cite{shin2021protein},  GVP \cite{jing2020learning} and Fold2Seq \cite{cao2021fold2seq}, respectively to generate the protein's CDR regions. For GVP and Fold2Seq, whose output is the entire protein sequence, we also run ANARCI to extract the CDRs. The output from all the models is then processed and filtered for the final property analysis.}
\label{fig:sys_overview}
\end{figure}

\section{Related Work}

The protein/antibody design problem has been recently addressed using multiple machine learning  approaches. For example, the sequence-based methods of \cite{alley2019unified, saka2021antibody, akbar2021silico} formulated the problem as an autoregressive process, leveraging deep recurrent networks such as an LSTM, to design the entire antibody sequence. On the other hand, the work of \cite{shin2021protein} focused on the variable complementarity determining regions (CDRs), which are complex and highly diverse parts of the sequence, determining the specificity of the antibody. Instead of recurrent networks, they applied causal dilated convolutions to model the autoregressive likelihood. 

Alternatively, ML-physics hybrid  methods 
have also been explored. For instance, \cite{tischer2020design}, proposed a system, that starts from a random protein sequence, and the search process is guided by the physics-based constraints to satisfy the desired structural motifs. EVOdesign \cite{Pearce2019EvoDesignDP}, on the other hand, is a structure-based approach that uses the evolutionary profiles to guide the sequence search. It first identifies the structural analogs from PDB and then constructs a structural sequence profile using MSA. The physics-based force fields are then used to search for the low free-energy sequence states. 

Recently, a class of deep generative models that account for the 3D structural constraints, have been proposed. For example, \cite{NEURIPS2019_f3a4ff48} used a generative model for protein sequences design given a target structure, represented as a graph over the residues. The key challenge was to account for long-range dependencies in the protein sequence that are usually short-range in the 3D structure space.

Based on \cite{NEURIPS2019_f3a4ff48} the work of \cite{jing2020learning} then proposed Geometric Vector Perceptrons (GVPs) to allow for the embedding of geometric information at nodes and edges without reducing such information to scalars that may not fully capture complex geometry. The proposed graph network ensures that the vector and scalar outputs are equivariant and invariant with respect to rotations and reflections.

Designing proteins based on a rigid backbone structure is known to restrict the diversity and novelty of the sequences. For this reason, \cite{cao2021fold2seq} proposed Fold2Seq, a transformer-based generative framework for designing protein sequences conditioned on a generic target fold, rather than the specific high resolution 3D structure. The fold here was defined as the spatial arrangement of the local secondary structure elements. The method also uses joint sequence–fold embedding to better capture the relationship between the two modalities.

Finally, a recent work of \cite{jin2021iterative} proposed a generative model to specifically design only the CDRs of antibodies rather than the whole protein sequence. The approach is based on co-designing the CDR sequence and 3D structure of CDR as graphs. The output is built autoregressively while iteratively refining its predicted global structure, which in turn guides the next residue choice. 

\begin{figure}[!t]
\centering
\includegraphics[width=0.9\textwidth]{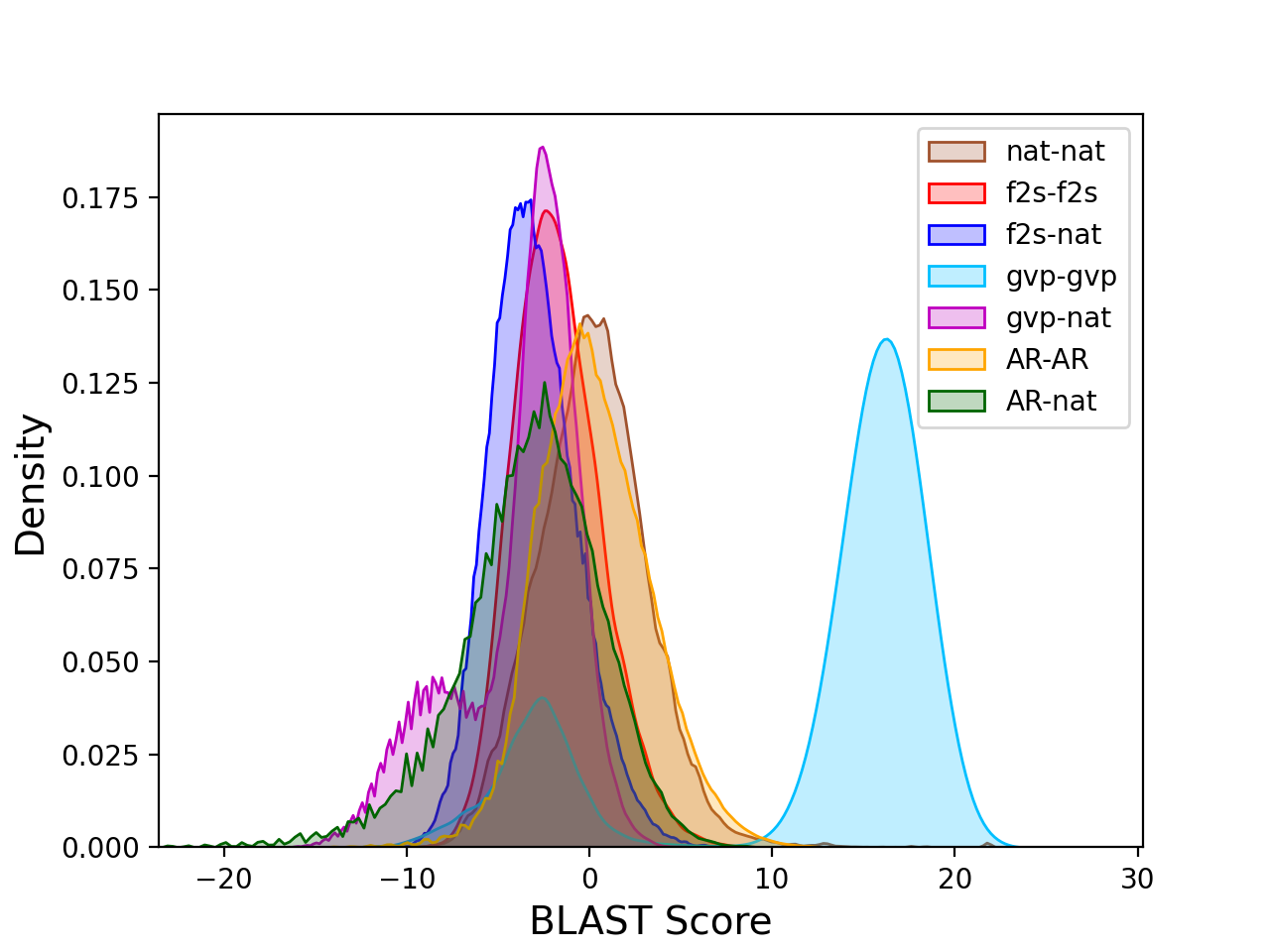}
\caption{Kernel density plot as a function of pairwise sequence similarity within the same ensemble (f2s= Fold2Seq, gvp=GVP, AR=AR, nat=Natural) and with the natural ensemble (*-nat). Higher score indicates better similarity.}
\label{fig:sim}
\end{figure}

\section{Method}

Figure \ref{fig:sys_overview} shows the overview of our benchmarking study.

\textbf{Models} We examine three different models. AR \cite{shin2021protein}: the autoregressive approach that uses the causal dilated convolutions for the input prefix sequence to generate the CDR3 protein subsequence. GVP \cite{jing2020learning}:  the encode-decoder GNN model that uses Geometric Vector Perceptrons to encode the scalar and vector structure information which is then decoded in the autoregressive manner to generate the entire protein sequence. Fold2seq \cite{cao2021fold2seq}: another encode-decoder model based on transformer architecture that embeds the fuzzy 3D protein structure information (fold) in the joint sequence-fold embedding space and then decodes it autoregressively into the corresponding protein sequence. For the GVP and Fold2Seq outputs we run ANARCI \cite{dunbar2016anarci} to extract the CDR regions. 

\textbf{Sequence Design}
The sequence and structure corresponding to the  Chain A of pdb id 3K3Q were used as inputs for this study. It is worth noting that this structure is included in the training of both GVP and Fold2Seq model, whereas a maximum of 58.94\% sequence identity was found to be present between the input sequence and the AR training set.

We compare the full sequences as well as the CDRs across the generated ensembles. For this purpose,
we extracted the CDRs from the generated sequences using the IMGT numbering scheme as returned by  ANARCI software \cite{dunbar2016anarci}.  For the extracted CDRs, we estimate the percentage of unique sequences (uniqueness). Sequences  that contain glycosylation sites, asparagine deamination motifs, or sulfur-containing amino acids (cysteine and methionine) were removed.  
For the AR model, we optionally considered an extra filter to exclude sequences that do not end with the final beta-strand of the nanobody template as in \cite{shin2021protein}. We denote the approach with final beta-strand filtering by \emph{AR filtered}, while we call \emph{AR unfiltered} the version without final beta-strand filtering. 

\textbf{Evaluation Metrics}
We define the set of the generated sequences (structures) conditioned on sequence/structure $j$ as $\mathcal{G}_j$. 
In structure-based design,  Sequence Recovery rate is defined as  ($SRR$) for $\bm{y}_j$ as $SRR_{structure}(j) = \frac{1}{|\mathcal{G}_j|}\sum_{g \in \mathcal{G}_j} SIM(\bm{x}_g, \bm{x}_j)$.
A global alignment scheme and BLAST62 matrix, with a gap opening penalty of -10 and gap extending penalty of -1, were used for estimating pairwise sequence identity (SIM) and  alignment score.
Negative log likelihood (NLL) was estimated  by using the autoregressive (AR) generative model trained on 1.2 million natural llama nanobody sequences \cite{shin2021protein},  as following:
$NLL= -\sum_{k=1}^{K}{\log(p(x_k|X_{<k}))}$, which is  sum of the cross-entropy between the true residue at each position and the predicted distribution over possible residues, conditioned on the preceding characters. 
 Structural recovery of the three sequence design models by predicting the 3D structure of the top 100 generated sequences using pretrained models from Alphafold2\footnote{https://github.com/kalininalab/alphafold\_non\_docker}.

\section{Results}



\begin{table}[!t]
\caption{Sequence recovery rate (SRR) and NLL from the autoregressive sequence model \cite{shin2021protein} trained on natural llama nanobody repertoire for the sequences generated by the comparison methods and the sequences from the natural llama library, synthetic library, and next-generation sequencing library.}

\begin{tabularx}{1.0\textwidth} { 
  | >{\raggedright\arraybackslash}X 
  | >{\centering\arraybackslash}X 
  | >{\centering\arraybackslash}X 
  | >{\raggedleft\arraybackslash}X | }

 \hline
 \textbf{Model} & \textbf{Seq Recovery Rate (\%)} & \textbf{NLL}\\
 \hline
 Fold2Seq &  30.711 & 2.572\\
 \hline
 GVP  &  40.131 & 2.987\\
 \hline
 AR &  48.865 & 0.375\\
\hline
Natural & -- & 0.371\\
\hline
Synthetic & -- & 4.912\\
\hline
NGS & -- & 5.102 \\
\hline
\end{tabularx}
\label{full}
\end{table}

\begin{table}[!h]
\caption{Uniqueness and novelty of the  CDR3, CDR2, and CDR1 regions of the sequences generated by Fold2Seq, GVP,  AR without final beta-stand filtering (AR unfiltered), AR with final beta-strand filtering (AR filtered),  and the sequences from the natural llama library. Note that the AR approach has been trained to generate CDR3 only, given the preceding portion of a ground truth sequence.}

\begin{tabular}{ll|c|c|c|c|c|}
\cline{3-7}
                                            &                & Fold2Seq & GVP   & AR unfiltered & AR filtered & Natural Llama \\ \hline
\multicolumn{1}{|l|}{\multirow{2}{*}{CDR3}} & Uniqueness     &  100        &  88.33 & 87.57         & 13.85       & 100         \\ \cline{2-7} 
\multicolumn{1}{|l|}{}                      & Novelty        &  43.36    &  52.71 & 11.92         & 8.97        & 52.64          \\ \cline{2-7} 
\hline\hline
\multicolumn{1}{|l|}{\multirow{2}{*}{CDR2}} & Uniqueness     &  100        &  9.15   &    --           &    --         &     100          \\ \cline{2-7} 
\multicolumn{1}{|l|}{}                      & Novelty        & 58.70    &  9.15    &         --      &    --         &        83.83       \\ \cline{2-7} 
 \hline\hline
\multicolumn{1}{|l|}{\multirow{2}{*}{CDR1}} & Uniqueness     &  92.49    &  56.20 &     --          &   --          &        100       \\ \cline{2-7} 
\multicolumn{1}{|l|}{}                      & Novelty        & 60.75    &  51.99 &     --          &     --        &      83.37             \\ \cline{2-7} 
\hline 
\end{tabular}
\label{novelty}
\end{table}

\begin{figure}
\centering
\includegraphics[width=1.0\textwidth]{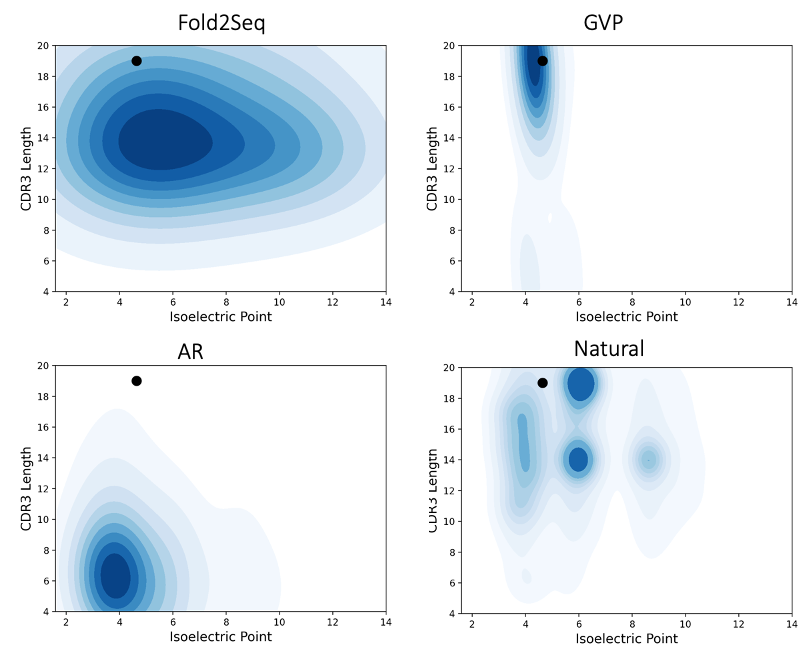}
\caption{2D kernel density plot as a function of isoelectric point and length of generated/natural CDR3 sequences. Black dot indicates the groundtruth CDR3.}
\label{fig:density}
\end{figure}

\begin{figure}[!hb]
    \centering
    \begin{minipage}[t]{.67\textwidth}
        \centering
        \includegraphics[width=\linewidth]{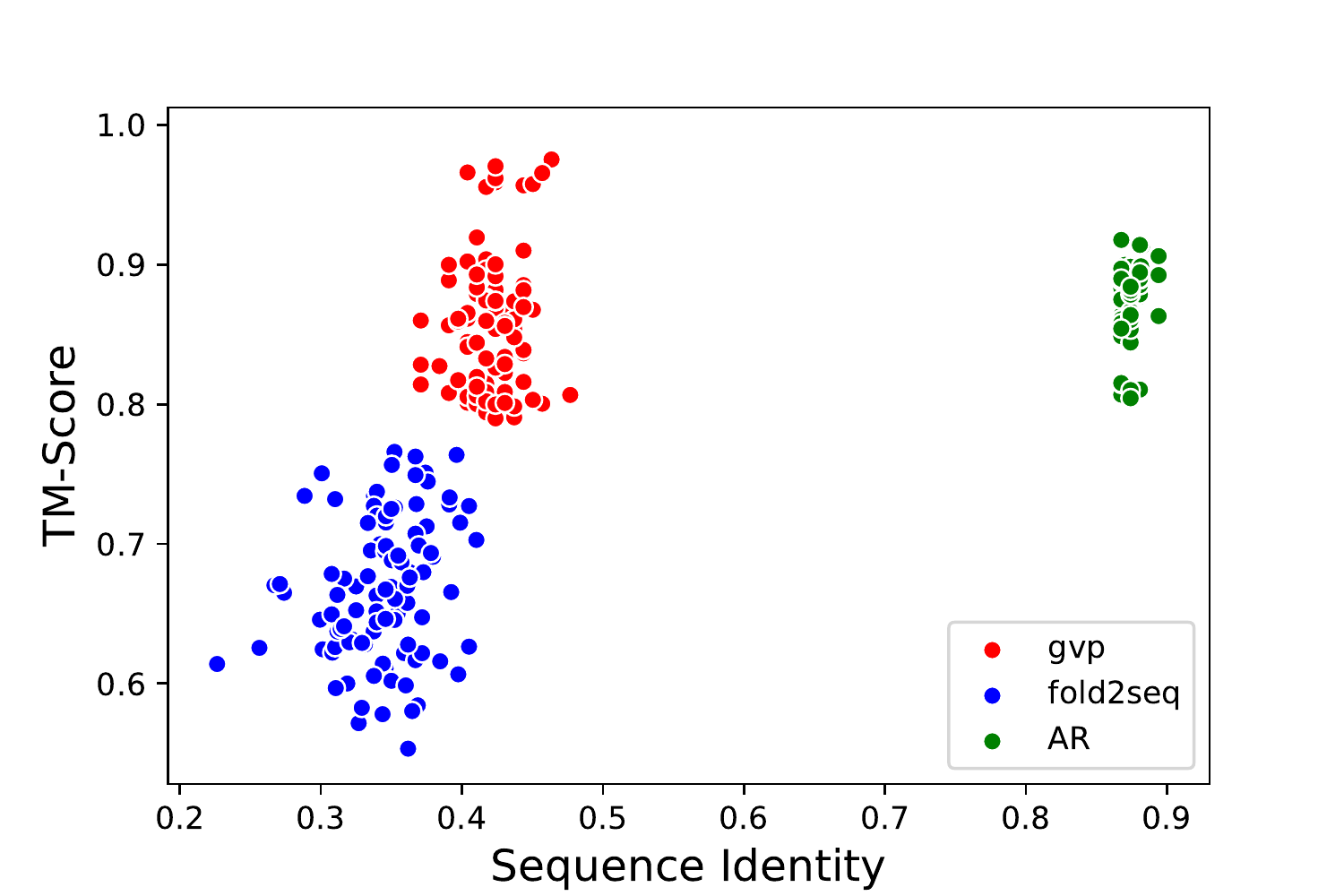}
    \end{minipage}%
    \begin{minipage}[t]{.43\textwidth}
        \centering
        \includegraphics[width=0.99\linewidth]{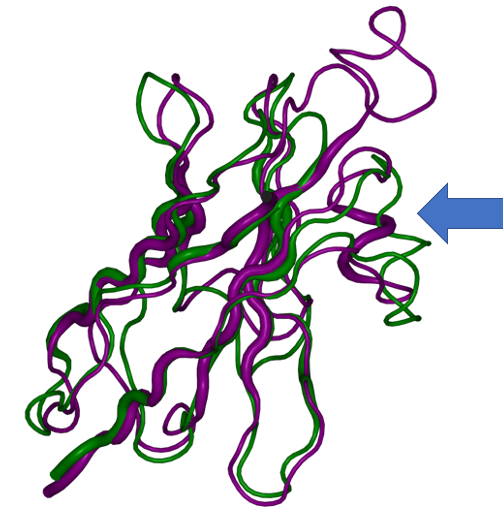}
    \end{minipage}
     \caption{(a) Scatter plot of sequence identity and structure similarity (TM-score) for Fold2Seq, GVP and AR generated sequences. The structure of the generated sequences are estimated using AlphaFold and the TM-Score is calculated with the ground truth structure using TM-Align. Note that since the sequence identity is computed over the entire protein chain, the AR method, generating only the CDR region (as opposed to others, that generate the whole sequence), by construction has higher similarity to the ground truth. (b) Alignment of the structure  corresponding to a Fold2seq generated sequence  (green) with the ground truth structure for PDB Id 3K3Q chain A using TM-align. Arrow indicates CDR3. TM-score is 0.75 and backbone RMSD is 1.86 \AA. }
     \label{fig:sub1}
\end{figure}

Table \ref{full} reports the average sequence recovery rate and average negative log likelihood from the trained autoregressive model in \cite{shin2021protein}, estimated using  10k generated sequences.  The autoregressive sequence model provides highest sequence recovery rate, followed by GVP and Fold2Seq. Both methods yield sequences that share $\geq$ 30\% identity on average with the sequence of the input structure, implying fold-consistency of the generated sequences, as   a  30\% sequence identity threshold typically suggests fold homology \cite{pearson2013introduction}. SRR of  GVP is higher than Fold2Seq,  consistent with earlier obeservation that a structure-based model yields higher sequence recovery than a fold-based model \cite{cao2021fold2seq}.  
The negative log likelihood (NLL) from the trained autoregressive model was found to be consistent with the experimentally reported thermostability of unseen llama nanobody sequences, which is an important aspect of  nanobody fitness \cite{shin2021protein}. Therefore, we also estimate the NLL of the generated sequences using Fold2Seq, GVP, AR, as well as a state-of-art synthetic library (Synthetic, constructed combinatorically using the position-specific amino acid frequencies of nanobody sequences with crystal structures in the PDB database) \cite{mcmahon2018yeast} and a large collection of nanobody sequences from next-generation sequencing repositories (ngs) \cite{deszynski2021indi}. As expected, AR-generated sequences show  a NLL close to the natural nanobodies that were used for training of the model. Both synthetic and ngs libraries show very high NLL, indicating the bias of the trained AR toward natural llama nanobody repertoire. Among the two 3D-conditional generative models, interestingly, F2S sequences show lower NLL than GVP, implying F2S sequences are closer to the broader llama nanobody sequences.

As reported in Table \ref{novelty}, we estimate uniqueness and novelty of all three generative models with respect to the training set  used for the AR model training. Fold2Seq outperforms both GVP and AR by a significant margin in term of uniqueness for all CDRs. Trend in novelty is not as clear though between GVP and Fold2Seq: while GVP produces more novel CDR3s, Fold2Seq performs better for CDR1s and CDR2s in term of novelty.  For reference we also include in Table \ref{novelty} uniqueness and novelty for 10,000 sequences selected at random from the natural llama library of \cite{shin2021protein}.   This result implies that a fuzzy representation of a 3D fold allows generating sequences with novel and unique CDRs, when compared to a sequence-based and a backbone structure-based model.

This result is confirmed as we estimate the pairwise sequence similarity within the generated ensemble from a particular model (Figure \ref{fig:sim}). When compared to the database comprised of natural llama nanobody sequences, Fold2Seq sequences are more diverse (low similarity score within the ensemble)  than GVP and AR sequences. GVP sequences show an interesting characteristic -- majority of sequences are highly similar - indicating a mode collapse.
In term of similarity with the natural CDR3s, Fold2Seq sequences are more distant, followed by GVP and AR. The similarity distribution corresponding to GVP seem narrower, consistent with lack of diversity in generated samples.

 Further, we analyze the physicochemical properties, such as isoelectric point (pI) and length of the CDR3 sequences. Figure \ref{fig:density} shows that the natural nanobody CDR3s populate broad ranges of isoelectric point and sequence length, with a preference toward pI around 6 and length around 14 and 18. Fold2Seq produces a significant coverage of the  natural sequences, with a clustering around pI = 6  and length = 14, while still exhibiting non-zero density around the input.  GVP produces sequences that are very close to the  one corresponding to the input pdb only, while  AR tends to generate short (around 6 amino acid) sequences with pI around 4.




Finally, we compare the structural recovery of the three generative frameworks  by predicting the 3D structure of  100 random generated sequences using  Alphafold2. Figure \ref{fig:sub1}(a) shows that,  almost all of the  sequences generated by the Fold2Seq and GVP share a sequence identity of $\geq$ 30\% and TM-score of $\geq$ 0.5 with the groundtruth,  indicating both methods perform similarly in term of recovering correct fold topology \cite{xu2010significant}. However,  GVP sequences exhibit higher TM-score than Fold2Seq ones, consistent with high consistency of the generated sequences with the input  while  lacking  coverage and diversity (see Figure \ref{fig:sub1}(a)). The AR sequences show high sequences identity as well as high TM-score, which is not surprising given that the model ``grafts'' a short (around 6 residue) sequences within the input sequence. In contrast, though Fold2Seq generates diverse and dissimilar sequences,  corresponding structures turn out to be high-consistency, with a 1.8-3 \AA~ 
backbone RMSD from the input structure (see Figure \ref{fig:sub1}(b)). 

In conclusion, this benchmarking study highlights key performance differences of three protein sequence design models on the nanobody sequence generation task. The backbone based GVP model provides high consistency with the input, while lacking the   diversity present within the llama nanobody sequences. A sequence-based model, on the other hand, captures a limited spectrum of the llama nanobody repertoire. The sequence design model that employs a fuzzy representation of the corresponding 3D fold as the input and generates sequences by leveraging representation learned by join training on 3D folds and 1D  sequences provides better diversity, while still providing sufficient consistency with the input sequence/structure and coverage of the llama nanobody repertoire.

Ultimately, the choice of method should be guided by the task at hand. If one wishes to generate sequences very close to an input sequence/structure, GVP might be more appropriate. If one wishes to graft specific regions such as a short CDR3 within an input sequence, then the AR approach would be better suited. If one wishes to generate diverse sequences with broader coverage that ``extrapolate'' to a greater extent from a given input sequence, while still maintaining consistency with the input, then Fold2Seq would be a method of choice. 
\bibliography{refs}

\begin{thebibliography}{10}

\bibitem{akbar2021silico}
Rahmad Akbar, Philippe~A Robert, C{\'e}dric~R Weber, Michael Widrich, Robert
  Frank, Milena Pavlovi{\'c}, Lonneke Scheffer, Maria Chernigovskaya, Igor
  Snapkov, and Andrei Slabodkin.
\newblock In silico proof of principle of machine learning-based antibody
  design at unconstrained scale.
\newblock {\em BioRXiV}, 2021.

\bibitem{alley2019unified}
Ethan~C Alley, Grigory Khimulya, Surojit Biswas, Mohammed AlQuraishi, and
  George~M Church.
\newblock Unified rational protein engineering with sequence-based deep
  representation learning.
\newblock {\em Nature methods}, 16(12):1315--1322, 2019.

\bibitem{bannas2017nanobodies}
Peter Bannas, Julia Hambach, and Friedrich Koch-Nolte.
\newblock Nanobodies and nanobody-based human heavy chain antibodies as
  antitumor therapeutics.
\newblock {\em Frontiers in immunology}, 8:1603, 2017.

\bibitem{cao2021fold2seq}
Yue Cao, Payel Das, Vijil Chenthamarakshan, Pin-Yu Chen, Igor Melnyk, and Yang
  Shen.
\newblock Fold2seq: A joint sequence (1d)-fold (3d) embedding-based generative
  model for protein design.
\newblock In {\em International Conference on Machine Learning}, pages
  1261--1271. PMLR, 2021.

\bibitem{deszynski2021indi}
Piotr Deszynski, Jakub Mlokosiewicz, Adam Volanakis, Igor Jaszczyszyn, Natalie
  Castellana, Stefano Bonissone, Rajkumar Ganesan, and Konrad Krawczyk.
\newblock Indi-integrated nanobody database for immunoinformatics.
\newblock {\em medRxiv}, 2021.

\bibitem{dunbar2016anarci}
James Dunbar and Charlotte~M Deane.
\newblock Anarci: antigen receptor numbering and receptor classification.
\newblock {\em Bioinformatics}, 32(2):298--300, 2016.

\bibitem{NEURIPS2019_f3a4ff48}
John Ingraham, Vikas Garg, Regina Barzilay, and Tommi Jaakkola.
\newblock Generative models for graph-based protein design.
\newblock In {\em Advances in Neural Information Processing Systems},
  volume~32. Curran Associates, Inc., 2019.

\bibitem{jin2021iterative}
Wengong Jin, Jeremy Wohlwend, Regina Barzilay, and Tommi Jaakkola.
\newblock Iterative refinement graph neural network for antibody
  sequence-structure co-design, 2021.

\bibitem{jing2020learning}
Bowen Jing, Stephan Eismann, Patricia Suriana, Raphael~JL Townshend, and Ron
  Dror.
\newblock Learning from protein structure with geometric vector perceptrons.
\newblock {\em arXiv preprint arXiv:2009.01411}, 2020.

\bibitem{mcmahon2018yeast}
Conor McMahon, Alexander~S Baier, Roberta Pascolutti, Marcin Wegrecki, Sanduo
  Zheng, Janice~X Ong, Sarah~C Erlandson, Daniel Hilger, S{\o}ren~GF Rasmussen,
  and Aaron~M Ring.
\newblock Yeast surface display platform for rapid discovery of
  conformationally selective nanobodies.
\newblock {\em Nature structural \& molecular biology}, 25(3):289--296, 2018.

\bibitem{Pearce2019EvoDesignDP}
Robin Pearce, Xiaoqiang Huang, Dani Setiawan, and Yang Zhang.
\newblock Evodesign: Designing protein-protein binding interactions using
  evolutionary interface profiles in conjunction with an optimized physical
  energy function.
\newblock {\em Journal of molecular biology}, 431 13:2467--2476, 2019.

\bibitem{pearson2013introduction}
William~R Pearson.
\newblock An introduction to sequence similarity (“homology”) searching.
\newblock {\em Current protocols in bioinformatics}, 42(1):3--1, 2013.

\bibitem{saka2021antibody}
Koichiro Saka, Taro Kakuzaki, Shoichi Metsugi, Daiki Kashiwagi, Kenji Yoshida,
  Manabu Wada, Hiroyuki Tsunoda, and Reiji Teramoto.
\newblock Antibody design using lstm based deep generative model from phage
  display library for affinity maturation.
\newblock {\em Scientific reports}, 11(1):1--13, 2021.

\bibitem{shin2021protein}
Jung-Eun Shin, Adam~J Riesselman, Aaron~W Kollasch, Conor McMahon, Elana Simon,
  Chris Sander, Aashish Manglik, Andrew~C Kruse, and Debora~S Marks.
\newblock Protein design and variant prediction using autoregressive generative
  models.
\newblock {\em Nature Communications}, 12, 2021.

\bibitem{tischer2020design}
Doug Tischer, Sidney Lisanza, Jue Wang, Runze Dong, Ivan Anishchenko, Lukas~F
  Milles, Sergey Ovchinnikov, and David Baker.
\newblock Design of proteins presenting discontinuous functional sites using
  deep learning.
\newblock {\em bioRxiv}, 2020.

\bibitem{xu2010significant}
Jinrui Xu and Yang Zhang.
\newblock How significant is a protein structure similarity with tm-score= 0.5?
\newblock {\em Bioinformatics}, 26(7):889--895, 2010.

\end{thebibliography}
\bibliographystyle{plain}

\end{document}